\documentclass{aa}
\usepackage{txfonts}
\usepackage{natbib}
\usepackage[dvipsnames]{xcolor}
\usepackage{hyperref}
\usepackage[normalem]{ulem}
\hypersetup{
    colorlinks=true,
    linkcolor={blue},
    citecolor={blue}
    }
\bibpunct{(}{)}{;}{a}{}{,}
\usepackage{orcidlink}
\usepackage[dvipsnames]{xcolor}

\newcommand{\dd}{\text{d}}
\newcommand{\mb}{\vec{B}}
\newcommand{\me}{\vec{E}}
\newcommand{\mv}{\vec{v}}
\newcommand{\mF}{\vec{F}}
\newcommand{\mR}{\vec{R}}

\newcommand{\pa}{v_\parallel}

\begin{document}

\title{Test particle sampling and particle acceleration in a 2D coronal plasmoid-mediated reconnecting current sheet}
 
\author{Eilif S. Øyre\inst{1}\inst{2}\orcidlink{0009-0005-6122-1987}
  \and Boris V. Gudiksen\inst{1}\inst{2}\orcidlink{0000-0003-0547-4902}
  \and Lyndsay Fletcher\inst{1}\inst{2}\inst{3}\orcidlink{0000-0001-9315-7899}
  } 

\offprints{Eilif. S. Øyre, \email{e.s.oyre@astro.uio.no}}

\institute{Rosseland Centre for Solar Physics, University of Oslo, Sem Sælands vei 13, Oslo, Norway 
\and Institute of Theoretical Astrophysics, University of Oslo, Sem Sælands vei 13, Oslo, Norway
\and Scottish Universities Physics Alliance, School of Physics and Astronomy, University of Glasgow, Glasgow, G12 8QQ United Kingdom}

\date{Received 7 May 2025 / Accepted 30 August 2025}

\abstract 
{Solar flares accelerate electrons, creating non-thermal energy distributions. However, the acceleration sites and dominant acceleration mechanisms remain largely unknown.} 
{We study the characteristics of electron acceleration and subsequent non-thermal energy distribution in a 2D coronal plasmoid-mediated reconnecting current sheet.}
{We used test particles and the guiding centre approximation to transport electrons in a static coronal 2D fan-spine topology magnetohydrodynamic (MHD) snapshot. The snapshot was from a Bifrost simulation that featured plasmoid-mediated reconnection at a current sheet. To sample initial particle conditions that lead to non-thermal energies, we used importance sampling. In this way, the characteristics of the non-thermal electrons were statistically representative of the MHD plasma.} 
{The energy distribution of the electrons forms a non-thermal power law that varies with our tolerance of the guiding centre approximation's validity, from no obvious power law to a power law with an exponent of -4 (the power law also depends on the statistical weighing of the electrons). The non-thermal electrons gain energy through a gradual betatron acceleration close to magnetic null points associated with plasmoids.}
{In this static, asymmetric, coronal, 2D fan-spine topology MHD configuration, non-thermal electron acceleration occurs only in the vicinity of null points associated with magnetic gradients and electric fields induced by plasmoid formation and ejection. However, the guiding centre approximation alone is not sufficient to properly estimate the shape of the non-thermal power law since, according to our results, electron acceleration is correlated with the adiabaticity of the particles' motion. The results also show that the particle power law formation is biased by the test particle sampling procedure.}

\titlerunning{Test particle sampling and particle acceleration in a 2D reconnecting current sheet}
\maketitle

\section{Introduction}\label{section:introdution}
Solar flares cannot be explained without considering the effects of non-thermal electrons. The presence of these electrons is inferred directly through X-ray and radio emission, and indirectly through the optical, ultraviolet, and extreme ultraviolet radiation produced by the collisional heating they cause \citep{benzFlareObservations2016}.  Magnetic reconnection is believed to be responsible for providing the energy that is required for the particles to reach non-thermal velocities, but the processes responsible for the particle acceleration are debated \citep{gordovskyyCombiningMHDKinetic2019, browningUnsolvedQuestionsFuture2024}. The observed signatures point towards a non-thermal power law energy distribution that varies significantly from flare to flare \citep{2008ApJ...677.1385c} and also as a function of space and time within a flare \citep{2004A&A...426.1093G,2006A&A...458..641G}, indicating that the particle acceleration is sensitive to the large-scale magnetic configuration.

Magnetohydrodynamic (MHD) simulations are able to resolve large scales realistically \citep{viggo}, and they are able to approximate regions with magnetic reconnection using sophisticated resistivity models (see \citealt{faerderComparativeStudyResistivity2023} and references therein). Nevertheless, kinetic models are needed to describe the acceleration and transport of non-thermal particles, since MHD assumes thermal energy distributions.

Non-thermal particle beam models allow us to transport already accelerated particles in MHD simulations and deposit their energy along the way. In 3D, this was first done by \cite{frognerAcceleratedParticleBeams2020a} and was followed by more realistic beam propagation in \cite{2024A&A...683A.195F}. In \cite{ruanFullySelfconsistentModel2020} and \cite{druettExploringSelfconsistent25D2024}, the energy depositions are fed back into the MHD simulation, and the beams' energy gains are subtracted from the MHD energy density. This enables self-consistent simulation of the energetics and atmospheric response of large flares where the accelerated particle beams contribute to chromospheric heating and upflows. However, particle beam models rely on the free parameters that define the initial beam velocity distribution. The study of particle acceleration helps constrain these parameters.

Non-thermal particle acceleration can occur on both large and small (kinetic) scales. Particle-in-cell (PIC) simulations by \cite{drakeElectronAccelerationContracting2006} have shown that repeated first-order Fermi acceleration in contracting magnetic islands (or plasmoids) can be an efficient mechanism. In addition, PIC simulations of the magnetosphere by \cite{egedalLargescaleElectronAcceleration2012} indicate that electrons in reconnection exhausts are significantly accelerated by large-scale parallel electric fields. However, PIC simulations are too computationally expensive to include both the kinetic and large scales of flares on the Sun. A compromise is to surround PIC simulations with MHD boundaries, as done by \cite{baumannKINETICMODELINGPARTICLE2013}. They found that parallel electric fields from the magnetic reconnection regions contributed the most to electron acceleration, and that the shape of the non-thermal power law depended mostly on the geometry and evolution of the magnetic field. Another approach is to simultaneously represent the electrons in the plasma as both a fluid and as particles. Doing this in 2D, while at the same time eliminating kinetic-scale parallel electric fields and approximating the particle motion with the guiding centre approximation (GCA), \cite{arnoldElectronAccelerationMacroscale2021} simulated strong non-thermal acceleration via the first-order Fermi process, similar to \cite{drakeElectronAccelerationContracting2006}, but this time with a larger simulation domain, of the order of $10^4$ km. Finally, a third approach is to combine large-scale MHD simulations with test particles, which have been shown to accelerate electrons up to MeV energies \citep{gordovskyyParticleAccelerationTransport2014} and can be used to synthesise observational signatures such as hard X-ray emission from MHD flare simulations \citep{pintoThermalNonthermalEmission2016}.

Test particles do not alter the external electromagnetic field in which they are embedded, nor do they interact with each other. This loss of self-consistency comes with the benefit of a reduction in computational cost, as the particle simulation reduces to a Monte Carlo problem and the external fields can be given by MHD or even analytically prescribed. The reduced cost makes it possible to simulate a very large number of particles, tracing them over very large distances and up to very high energies. From particle trajectories, one can study the characteristics of acceleration and infer statistical properties of the plasma \citep{marchandTestparticleSimulationSpace2010}. Stochastic approaches that simulate the interaction of test and background particles have frequently been used, starting with \cite{1991A&A...251..693M} and explored by \cite{fletcherGenerationLooptopImpulsive1995}, \cite{burgeEffectBinaryCollisions2014}, \cite{pintoThermalNonthermalEmission2016}, and \cite{borissovParticleAccelerationAnomalous2020}. Without background collisions, the test particle simulations rely on two strong assumptions: collisionless motion and negligible particle feedback.

Collisionless motion is a valid approximation as long as the paths of the particles are shorter than the mean free path of the plasma. The non-thermal particles are believed to be accelerated in the thin and hot solar corona \citep{kruckerHardXrayEmission2008}, making collisionless motion a common assumption in test particle studies \citep{rosdahlTestParticleAcceleration2010, gordovskyyForwardModelingParticle2020, zhaoMagneticIslandMerging2021, pallister2021, bacchiniParticleTrappingAcceleration2024, 2025MNRAS.536..609M}. Excluding particle feedback has been suggested to artificially increase particle acceleration as it ignores the effect of return currents \citep{rosdahlTestParticleAcceleration2010}. The exclusion is valid only as long as the electric current produced by the accelerated particles is negligible compared to the background currents. This depends on the fraction of non-thermal particles relative to the fraction of thermal particles. In some large flares, X-ray, microwave, and extreme ultraviolet measurements indicate that a large fraction, possibly all, of the electrons in a coronal volume can be accelerated \citep{2010ApJ...714.1108K,2014ApJ...780..107K}, but statistical studies show that in smaller flares the total energy in non-thermal particles is a smaller fraction of the peak thermal energy \citep{2020A&A...644A.172W} than in larger flares. From this we can infer that smaller flares accelerate a smaller portion of the particle population. This may restrict the use of test particles to smaller flares.

A source of smaller flares is magnetic reconnection in coronal, fan-spine topologies. Here, the current sheet at the magnetic null point can be a source of particle acceleration \citep{pallister2021}. If the current sheet is stretched thin with sufficient resolution and appropriate plasma resistivity, the plasmoid instability will occur \citep{furthFiniteResistivityInstabilitiesSheet1963}. Plasmoids of various sizes grow and merge with the exhaust of reconnection jets. This null point region resembles the flare model in \cite{nishizukaFermiAccelerationPlasmoids2013}, where plasmoid ejection and collision from the reconnection region drive particle acceleration.

We used a static snapshot from an MHD simulation of plasmoid-mediated reconnection in a 2D fan-spine magnetic topology as a background for test particle simulations, and studied the characteristics of emerging non-thermal electrons. The MHD simulation is of high enough resolution and large enough extent to resolve both plasmoid formation at the reconnecting null point and the large-scale fan-spine magnetic topology. We used the GCA for the particles' equations of motion and simulated particle ensembles that are statistically representative of the MHD plasma. To ensure the ensembles contain non-thermal electrons, we used importance weighing to bias the particle sampling with knowledge gained from an initial test particle simulation.

\section{Methods}\label{section:methods}
The test particle simulation consisted of four parts: the equations of motion of the particles; the electric and magnetic fields that the forces acting on the particles depend on; a method for choosing the initial conditions of the particles; and a method for solving the equations of motion.

\subsection{The particles' equations of motion}
A non-relativistic test particle embedded in an external magnetic ($\mb$) and electric field ($\me$) is subject to the Lorentz force,
\begin{equation}\label{eq:lorentzforce}
    \mF = q(\me + \mv \times \mb),
\end{equation}
where $q$ and $\mv$ are the charge and velocity of the particle, respectively.  In typical solar coronal conditions, electrons have a very small Larmor radius and fast gyration due to their low mass-to-charge ratio and due to strong magnetic fields. Resolving their helical motion quickly becomes a computationally demanding task. To overcome this problem, we used the GCA, which averages the particle's motion over its gyration. The set of equations that describe the motion of the guiding centre is valid as long as the electromagnetic field is sufficiently smooth over a gyration period. We considered the approximation valid whenever
\begin{equation}
    \frac{r_L}{L_B} \ll 1,
\end{equation}
where $r_L$ is the Larmor radius of the particle and $L_B = B/|\nabla B|$ is the characteristic length of the external magnetic field.

\cite{northropGuidingCenterApproximation1961} derived the equations describing the time evolution of the position of the guiding centre $\mR$ (in three dimensions) and the time evolution of its velocity component parallel to the magnetic field $\pa$. Essentially, the motion of the guiding centre is split into a component parallel to the magnetic field and a component perpendicular to the magnetic field. The perpendicular component is represented by a set of drifts.  Here we used the non-relativistic version of the equations (see \citealt{ripperdaComprehensiveComparisonRelativistic2018} for an extensive discussion of the equations). They are
\begin{subequations}\label{eq:eom}
\begin{align}
    \frac{\text{d}\vec{R}}{\text{d}t} &= v_\parallel\vec{b} +
        \vec{v}_E +  \vec{v}_{\nabla B} + \vec{v}_C + \vec{v}_{P} 
        \label{eq:gcadrifts}\\
    \frac{\text{d} v_\parallel}{\text{d} t} &= \frac{1}{m}\left(qE_\parallel -
        \mu\vec{b}\cdot\nabla B\right) + \vec{v}_E 
        \cdot \frac{\dd \vec{b}}{\dd t}, \label{eq:gcaparallel}
\end{align}
\end{subequations}
where $\vec{b} = \mb/B$ is the magnetic field direction, $E_\parallel = \vec{E}\cdot \vec{b}$ is the electric field component parallel to $\vec{b}$, $\mu$ is the magnetic moment of the particle (which is assumed to be constant in the GCA), and $m$ is the particle mass. The perpendicular drifts are
\begin{subequations}
\begin{align}
        \vec{v}_{E} &= \frac{\vec{E} \times \vec{b}}{B}
        &&\text{(}E\text{ cross }B\text{ drift),} \label{eq:exbdrift}\\
        \vec{v}_{\nabla B} &= \frac{\mu}{qB}(\vec{b}\times \nabla B) 
        &&\text{(gradient }B\text{ drift),}\label{eq:gradbdrift}        \\
        \vec{v}_C &= \frac{mv_\parallel}{qB}\left(\vec{b} \times
        \frac{\text{d}\vec{b}}{\text{d} t}\right)
        &&\text{(curvature drift),}\label{eq:curvaturedrift}\\
        \vec{v}_P &= \frac{m}{qB}\left(\vec{b} \times
        \frac{\text{d}\vec{v}_E}{\text{d} t}\right)
        &&\text{(polarisation drift).}\label{eq:polarisationdrift}
\end{align}
\end{subequations}
The first term in Eq. \eqref{eq:gcaparallel} is the acceleration by parallel electric fields. The second term is responsible for magnetic mirroring, and the third term is first-order Fermi acceleration. The latter can be associated with the energy change due to a curvature drift along the electric field, since
\begin{equation}\label{eq:fermi}
    \vec{v}_E \cdot \frac{\dd \vec{b}}{\dd t} = \frac{\vec{v}_C \cdot q\vec{E}}{mv_\parallel}.
\end{equation}

We defined the kinetic energy of the particle as the sum of three components, the parallel velocity, the speed of the gyration ($v_\perp$), and the magnitude of the vector sum of the drifts ($v_\text{Drifts}$):
\begin{equation}\label{eq:energy}
    K = \frac{m}{2} \left(v_\parallel^2 + v_\text{Drifts}^2 \right) + B\mu.
\end{equation}
When we evaluated the material derivatives in Eq. \eqref{eq:eom}, we neglected drifts other than $\vec{v}_E$, which is usually the dominant one. The material derivative becomes
\begin{equation}\label{eq:materialderivative}
\frac{\dd}{\dd t} \equiv \frac{\partial}{\partial t} + v_\parallel\vec{b} \cdot \nabla + \vec{v}_E \cdot \nabla,
\end{equation}
and we can then write the change of energy in the gyration as
\begin{equation}\label{eq:gyrationenergy}
    \frac{\dd (B\mu)}{\dd t} = \mu\left(\frac{\partial B}{\partial t} + v_\parallel\vec{b} \cdot \nabla B + \vec{v}_E \cdot \nabla B\right).
\end{equation}
Here, the second term is again associated with magnetic mirroring and does not alter the total energy. The first and third terms represent betatron acceleration. This is a total energy increase or decrease as a result of an increasing or decreasing magnetic field strength, either through explicit time variance of the magnetic field or due to a gradient perpendicular to the magnetic field direction. Betatron acceleration may be associated with the gradient $B$ drift since
\begin{equation}\label{eq:betatron}
    \mu\vec{v}_E \cdot \nabla B = \vec{v}_{\nabla B} \cdot q\vec{E}.
\end{equation}

\subsection{The electromagnetic field and its gradients}
In the experiments described in this section, we assumed the timescale of the test particle motion to be much shorter than the characteristic timescales of the electric and magnetic fields. That is, $\me$ and $\mb$ were assumed static, and the explicit time derivative in Eq. \eqref{eq:materialderivative} becomes zero.
We obtained the static electromagnetic field from a snapshot of an MHD simulation produced by the Bifrost code \citep{gudiksenStellarAtmosphereSimulation2011}. Bifrost solves the radiative MHD equations using a sixth-order, finite difference differential operator. The operator stencil is asymmetric so that the derivatives become staggered. To minimise the number of interpolations (which are of fifth order), Bifrost places its variables on a staggered grid. Bifrost is used to simulate the solar atmosphere realistically and with high performance and precision, all the way from the corona down to the photosphere, sometimes also including parts of the convection zone.

We selected a snapshot from the Bifrost simulation by \cite{faerderComparativeStudyResistivity2024} that uses hyper-diffusion to model resistivity. This time series simulates a 2D slice of an idealised coronal fan-spine magnetic configuration. There is no gravity nor radiative cooling or heating.  The initial magnetic field has a 70 $\mu$T negative parasitic polarity (at the lower boundary) embedded in a 30 $\mu$T positive background field (in the $z$-direction). The initial magnetic field and plasma velocity have no $y$ components and thus always lie in the $xz$ plane. Consequently, there are no electric field components parallel to the magnetic field since the electric current and field will always lie in the $y$-direction. The plasma is treated as an ideal, ionised solar abundance gas, with a mean molecular weight of 0.616 amu. The initial plasma density and temperature are homogeneous with values $3 \cdot 10^{-13}$ kg m$^{-3}$ and 0.61 MK, respectively. At the lower boundary, a constant horizontal velocity is imposed that has a peak value at the location of the inner spine of the magnetic field, advecting the inner spine parallel to the lower boundary. This stretches the current sheet at the null point, which eventually becomes so thin and long that the plasmoid instability is reached. From halfway in the time-evolving simulation, at a point where two plasmoids are present, we increased the resolution using linear interpolation, and ran the simulation over a small time span. We took the electromagnetic field and electron density from the final snapshot of this rerun to use as static fields in the test particle simulations.

Having the electromagnetic field on a grid, we needed interpolation to get the field values at an arbitrary position in space. For this purpose, we used cubic spline interpolation. We also needed to calculate the spatial derivatives $\nabla B$, $\nabla \vec{b}$, and $\nabla \vec{v}_E$. To do this, we used forward mode automatic differentiation locally, at each particle position, using the Julia package ForwardDiff \citep{revelsForwardModeAutomaticDifferentiation2016}. Having methods for evaluating the electromagnetic field and its gradients at arbitrary positions, we only needed the particle initial conditions to start simulating their trajectories.

\subsection{Particle sampling}
The method for choosing the initial conditions is crucial for finding non-thermal particles and inferring properties of the plasma. We wanted to study the statistical nature of the non-thermal particles (e.g. the probability of a particle becoming non-thermal from a specific set of initial conditions), so we needed the initial conditions of the test particles to be a statistically accurate representation of the particles in the solar atmospheric plasma. In our case, the particles needed to reflect the electron number density and the plasma temperature.

\subsubsection{Sampling initial positions and velocities}
We sampled the initial positions from the electron density distribution using rejection sampling. The rejection sampling algorithm is described in Appendix \ref{appendix:rejectionsampling}. The initial velocity was sampled from a Maxwell-Boltzmann distribution, where the distribution's temperature corresponds to the MHD temperature at the initial position of the particle. To obtain the Maxwell-Boltzmann distribution, we sampled each of the three velocity components independently from a Gaussian distribution. Finally, from the initial velocity components ($\vec{v}$) and the initial position ($\vec{r}$), we evaluated the initial guiding centre position, the initial parallel velocity, the initial gyration velocity, and the magnetic moment according to the formulas:
\begin{subequations}
\begin{align}
    \vec{R} &= \vec{r} + \frac{m}{qB(\vec{r})}\bigg\{\big[\vec{v} - \vec{v}_E(\vec{r}) \big] \times \vec{b}(\vec{r}) \bigg\}, \\
    v_\parallel &= \vec{v} \cdot \vec{b}(\vec{r}), \\
    \vec{v}_\perp &= \vec{v} - v_\parallel\vec{b}(\vec{r}) - \vec{v}_E(\vec{r}), \\
    \mu &= \frac{mv_\perp^2}{2B(\vec{r})}.
\end{align}
\end{subequations}
Here we have assumed that the perpendicular drift at $\vec{r}$ is dominated by $\vec{v}_E$.

\subsubsection{Re-sampling initial positions}
In the context of the entire simulation domain, non-thermal electrons are rare. It is very unlikely that an electron sampled from the electron density distribution will become non-thermal. This will lead us to simulate a very large number of `uninteresting' electrons. To deal with this problem, we used the results of an initial, unbiased test particle run as an indication of where acceleration occurs. Specifically, we used the relative energy gain as a function of the initial position to create a probability density distribution from which we drew the initial positions of a second run. This re-sampling will be biased towards sites that accelerated particles in the first ensemble, and consequently increase the number of accelerated particles in the second ensemble. To maintain the statistical representation of the plasma, the particles in the second run are weighed by their importance weights (see Appendix \ref{appendix:importancesampling} for details on importance sampling). The importance weight is defined as
\begin{equation}
    w(\vec{r}) = \frac{f(\vec{r})}{g(\vec{r})},
\end{equation}
where $\vec{r}$ is the particle initial position, $f$ is the electron density distribution, and $g$ is the probability density distribution from which we drew the initial positions in the re-sampling.

\subsection{Solving the equations of motion}
To solve Eqs. \eqref{eq:gcadrifts} and \eqref{eq:gcaparallel} we used the Julia package DifferentialEquations \citep{rackauckas2017}. Specifically, we use a second and third order L-stable Rosenbrock-W method, with adaptive time-stepping. All particles were solved for the same time span, but they were stopped if they exited the simulation domain, or if the GCA was considered invalid. The latter was determined by evaluating the criterion 
\begin{equation}\label{eq:tolerance}
    r_L/L_B < \text{tolerance},
\end{equation}
which we call the GCA tolerance. The criterion was evaluated at each integration time step. If the ratio is larger than the tolerance, the particle was stopped. To investigate the effects of the tolerance value, we ran the test particle ensemble multiple times with varying tolerance.

\section{Results}\label{section:results}
\subsection{The MHD snapshot} 
\begin{figure*}
    \centering
    \includegraphics[width=1\linewidth]{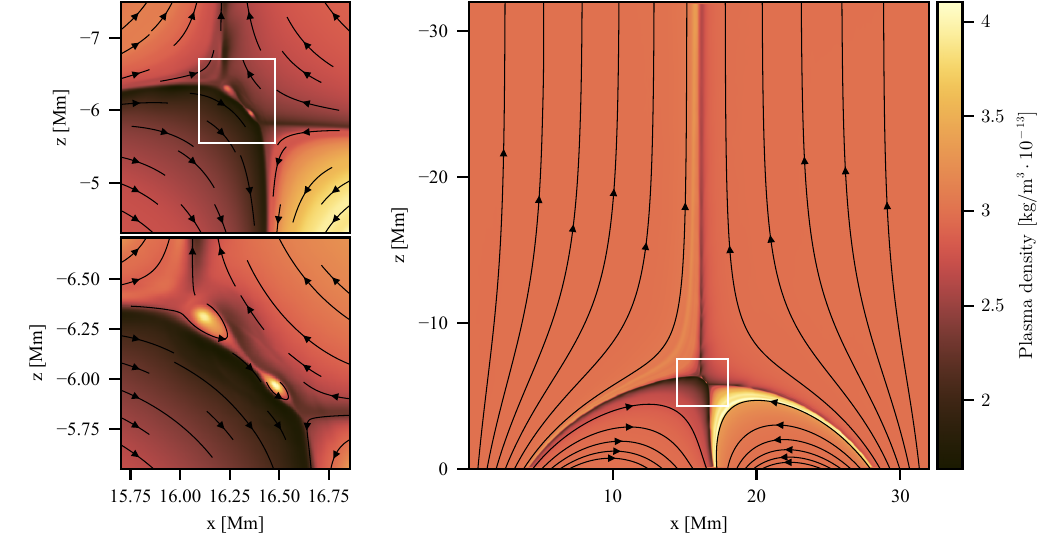}
    \caption{Plasma density of the MHD snapshot where we embedded test particles. The black streamlines show the magnetic field direction, and the two left panels shows the reconnection region in greater detail. The grid size is $8192^2$, and the resolution is 3.9 km.}
    \label{fig:mhdsnapshot}
\end{figure*}

The MHD snapshot is shown in Fig. \ref{fig:mhdsnapshot}. Here, the plasma density gradients illuminate the separatrices of the magnetic configuration. The 2D fan and inner spine are
merging with the outer spine at the null point, where two plasmoids are visible as high-density disks and with closed magnetic field lines. The upper-left plasmoid is on its way out into the upper-left reconnection exhaust, whereas the lower-right plasmoid is growing in size while moving towards the opposite exhaust.

\subsection{Test particle simulation}
\begin{figure}
    \centering
    \includegraphics[width=1.0\linewidth]{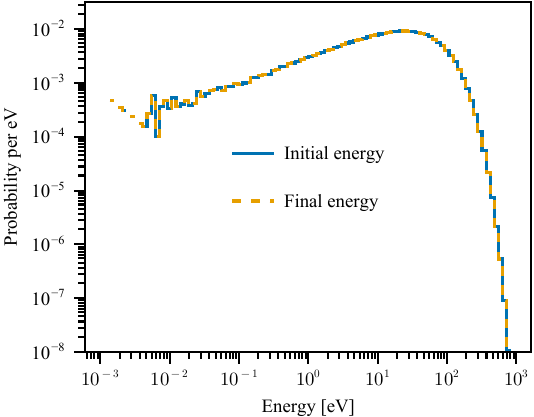}
    \caption{Normalised energy distribution of 10 million electrons before and after the first test particle simulation. The initial positions were sampled from the electron density of the MHD snapshot in Fig. \ref{fig:mhdsnapshot}. The initial velocities were sampled from a Maxwell-Boltzmann distribution at their initial temperature. The non-thermal electrons are too rare to be in the sample of this initial ensemble.}
    \label{fig:energydistr}
\end{figure}

We first simulated an ensemble of 10 million electrons for a duration of 1 second. The initial positions of the particles were distributed proportional to the electron density in the whole domain of the MHD snapshot in Fig. \ref{fig:mhdsnapshot}. Figure \ref{fig:energydistr} shows the energy distribution of all electrons before and after the simulation. The two distributions are hardly distinguishable.

\begin{figure}
    \centering
    \includegraphics[width=1\linewidth]{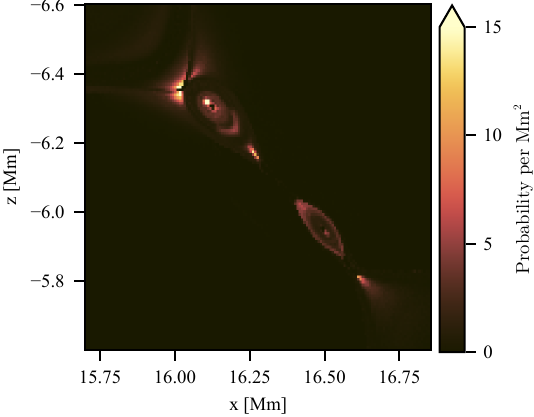}
    \caption{Median relative energy gain as a function of initial position of the third electron ensemble. The function is normalised so that it can be used as a probability density distribution. The function is practically zero outside the reconnection region. Values above 15 all have the same colour.}
    \label{fig:proposaldist}
\end{figure}

\begin{figure}
    \centering
    \includegraphics[width=1\linewidth]{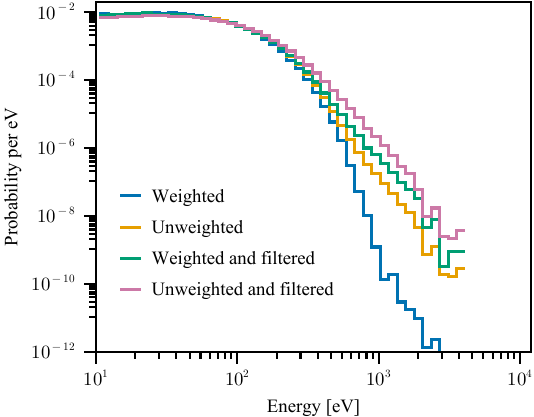}
    \caption{Normalised energy distribution after the third re-sampling of 13 million electrons. In this ensemble, the initial positions were sampled using Fig. \ref{fig:proposaldist} as a proposal distribution. The figure shows the final energy distribution with and without statistical weighting to account for the bias introduced by sampling from the proposal distribution, instead of the electron density. It also shows the distributions after applying a filter that excludes electrons outside the acceleration region, defined as the region displayed by the upper panel in Fig. \ref{fig:nonthermalpaths}. The distribution for energies lower than 10 eV is not shown.}
    \label{fig:energydistr-re-sample}
\end{figure}

\begin{figure}
    \centering
    \includegraphics[width=1\linewidth]{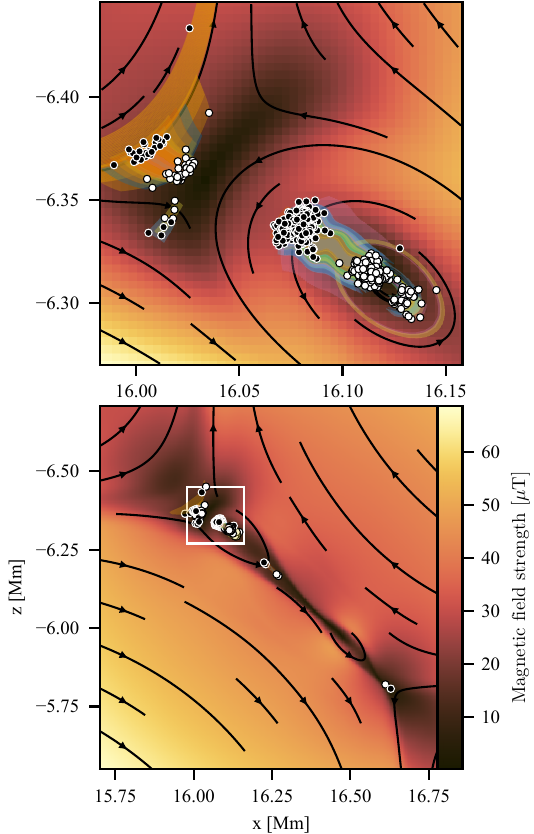}
    \caption{Magnetic field strength and field line segments of the current sheet overplotted with the trajectories of the 608 electrons with a final energy greater than 1 keV (excluding any particles with an initial energy above 1 keV). Their initial positions are shown as white dots, and the final positions as black dots.}
    \label{fig:nonthermalpaths}
\end{figure}

To increase the chances of sampling non-thermal electrons, we re-sampled initial conditions and ran a second ensemble of 3 million electrons. The new initial conditions were sampled from a proposal distribution constructed from the relative energy gain as a function of initial position of the first ensemble. This re-sampling procedure was repeated two more times, with an ensemble size of 5 million and 13 million, respectively. The third and final proposal distribution is displayed in Fig. \ref{fig:proposaldist}.
Figure \ref{fig:energydistr-re-sample} shows the initial and final electron energy distribution after simulating the third re-sampled electron ensemble. The final energy distribution is shown both with and without statistical weighing. The bins of the unweighted distribution have a height that corresponds to the number of electrons that fall within that bin, while the bins in the weighted distribution have a height that corresponds to the sum of the weights of the electrons that fall within that bin. The final energy distributions are also shown with and without a filter that excludes all particles not in the acceleration region, defined as the rectangular region displayed by the upper panel in Fig. \ref{fig:nonthermalpaths}.
In contrast to Fig. \ref{fig:energydistr}, Fig. \ref{fig:energydistr-re-sample} shows clear non-thermal power laws.

\begin{figure}
    \centering
    \includegraphics[width=1\linewidth]{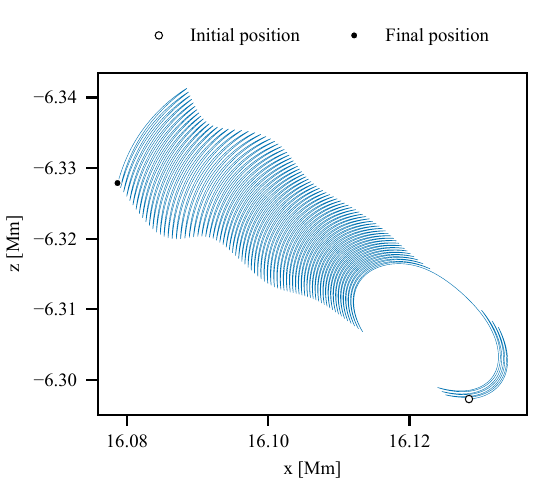}
    \caption{Trajectory of a betatron-accelerated non-thermal electron. The electron is initially upstream of the O-type null point, $\vec{E}\times\vec{B}$ drifting towards it. When it arrives downstream of the O-point, it drifts away from it.
    }
    \label{fig:trajectory}
\end{figure}

\begin{figure}
    \centering
    \includegraphics[width=1\linewidth]{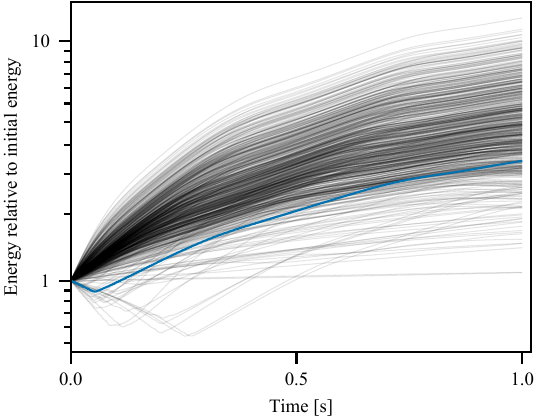}
    \caption{Time evolution of the kinetic energy of the 608 non-thermal electrons in the final test particle ensemble. The highlighted line corresponds to the trajectory in Fig. \ref{fig:trajectory}.}
    \label{fig:timescale}
\end{figure}

\begin{figure}
    \centering
    \includegraphics[width=1\linewidth]{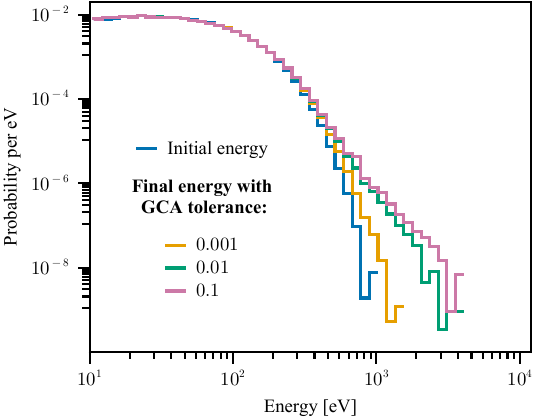}
    \caption{Initial and final energy distribution with a varying tolerance on the GCA's validity. All distributions are statistically weighted, and electrons outside the acceleration region are excluded. The distribution for energies lower than 10 eV is not shown.}
    \label{fig:tolerancechange}
\end{figure}

From the final energy distribution, we selected the accelerated electrons with a final energy greater than $1$ keV and plotted their trajectories in Fig. \ref{fig:nonthermalpaths}. The initial positions are mostly clustered into two groups, located around an O-type and an X-type magnetic null point associated with the upper-left plasmoid of Fig. \ref{fig:mhdsnapshot}.
Figure \ref{fig:trajectory} plots the trajectory of one of these electrons in detail, showing the characteristic path.
Its energy evolution is highlighted in Fig. \ref{fig:timescale}, which shows the energy evolution of all non-thermal electrons. The electrons have a gradual increase in energy and many of them have plateaued or even decreased in energy before 1\,s.
The GCA tolerance used to produce the results in Figs. \ref{fig:energydistr-re-sample}-\ref{fig:trajectory}, was $10^{-2}$. We re-ran the electron ensemble with a lower ($10^{-3}$) and a higher tolerance ($10^{-1}$), and their respective final energy distributions are shown in Fig. \ref{fig:tolerancechange}.

\section{Discussion}\label{section:discussion}
\subsection{Non-thermal power law formation}
The test particle simulations show sufficient electron acceleration to form a power law. However, we needed careful statistical sampling of initial positions to find high energy electrons. Out of the $\sim$$10^7$ electrons simulated, only $\sim$$10^3$ became non-thermal. In addition, the non-thermal particles have statistical weights much smaller than one, so when we in Fig. \ref{fig:energydistr-re-sample} look at the energy distribution of all electrons (not filtered), the weighted distribution differs significantly from the unweighted distribution. On the other hand, if we filter out the electrons outside the region around the acceleration sites, we get a significant power law even in the weighted distribution. The difference in the filtered distributions is lower because the non-thermal electrons are more common in the acceleration region, and because the sampling and electron densities are more uniform. The non-thermal tail of the weighted and filtered energy distribution carries an energy fraction of approximately $2 \cdot 10^{-4}$, which increases to $6 \cdot 10^{-4}$ for the corresponding unweighted distribution. For the unfiltered distributions, the total energy fractions are $10^{-7}$ and $10^{-5}$, respectively. Here we defined the non-thermal tail as the distribution above $1$ keV. A higher number of non-thermal electrons would increase the precision of the power law estimate, but the estimate would not necessarily be more accurate, since the power law formation is also heavily dependent on the GCA tolerance.

\begin{figure}
    \centering
    \includegraphics[width=1\linewidth]{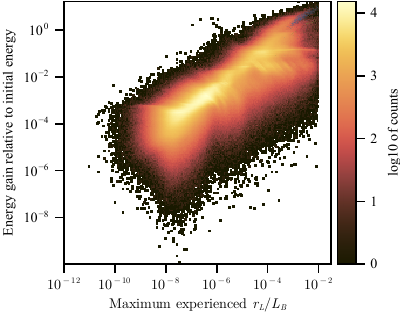}
    \caption{Correlation of an electron's relative energy gain with its highest experienced value of $r_L/L_B$. In this ensemble, the GCA tolerance was $10^{-2}$, hence the abrupt cutoff at $r_L/L_B = 10^{-2}$. The colour maps the counts on a logarithmic scale. The non-thermal electrons in the cluster around the O-type null point (of Fig. \ref{fig:nonthermalpaths}) are shown in blue. They all fall within the diagonal strand in the upper-right part of the histogram, where the relative energy gain is above 1 and $r_L/L_B$ is above $10^{-4}$. The non-thermal electrons in the cluster around the X-type null point are shown in green and fall within the more diffuse diagonal strand below.}
    \label{fig:breakdowncorrelation}
\end{figure}

The fact is that all the non-thermal electrons emerge from areas where the GCA is questionable, around magnetic null points. Figure \ref{fig:tolerancechange} shows that the power law hardens with increasing GCA tolerance, from no obvious power law to a power law with exponent -4. This suggests that the GCA is not sufficient to describe the motion of all potentially non-thermal electrons in this MHD configuration. To obtain a converging non-thermal power law, we would need to solve Eq. \eqref{eq:lorentzforce} directly in the areas of high $r_L/L_B$, similar to the hybrid schemes employed by \cite{birnElectronAccelerationDynamic2004}, \cite{stanierMagneticReconnectionParticle2013}, \cite{bacchiniCoupledGuidingCenter2020}, and \cite{pallister2021}. The need for full particle motion is emphasised by the positive correlation in Fig. \ref{fig:breakdowncorrelation}, where we have binned the final electron energies against the maximum $r_L/L_B$ value they experienced during their orbit. This figure indicates that the less valid the GCA is, the more likely the electron is to be accelerated. For the non-thermal electrons in Fig. \ref{fig:nonthermalpaths}, the GCA is least valid when they are closest to the null points. This is a result of $r_L/L_B$ being proportional to $B^{-2}$.

\subsection{Acceleration mechanisms}
The non-thermal electrons of Fig. \ref{fig:nonthermalpaths} drift away from their initial positions close to the magnetic null points due to the $\vec{E}\times\vec{B}$ drift. At the same time they are mirrored back and forth due to parallel gradients in the magnetic field strength. This is similar to the bouncing particles in the turbulent coronal loop-top simulation by \cite{bacchiniParticleTrappingAcceleration2024} where the electrons are trapped by magnetic mirroring and the mirroring represents the dominating parallel acceleration. However, the total energy increase is caused by betatron acceleration as the magnetic field strengthens when the electrons drift away from the null point, making Eq. \eqref{eq:betatron} positive. The mirroring acceleration itself does not alter the total energy. The electrons with a trajectory similar to the one in Fig. \ref{fig:trajectory} all start upstream of the O-type null point of the largest plasmoid. Here, the electric field points out of the plane, so that the electrons initially drift towards the O-point. When the magnetic field strength is too weak for magnetic mirroring, they orbit to the other side of the null point where the electric field points into the plane. From this point, their trajectories proceed like the electrons initialised downstream of the O-type null point, drifting away and betatron-accelerated.

There are no parallel electric fields, so the first term of Eq. \eqref{eq:gcaparallel} is always zero, but there is curvature in the magnetic field direction along the electron path, making the third term of Eq. \eqref{eq:gcaparallel} (the first-order Fermi acceleration term) non-zero. The energy change from this curvature acceleration is positive for the upper-left electron cluster in the upper panel of Fig. \ref{fig:nonthermalpaths} because here the curvature is parallel to $\vec{v}_E$, making Eq. \eqref{eq:fermi} positive. It is negative for the lower-right cluster, where the curvature is anti-parallel to $\vec{v}_E$. Both clusters experience a gradual increase in total energy, indicating that betatron acceleration dominates over the first-order Fermi acceleration. The plateauing of the energy evolution that can be seen in Fig. \ref{fig:timescale} can be explained by the fact that $\vec{v}_E \cdot \nabla B$ gradually reduces and changes sign in the middle between the two null points. The reduction in $\vec{v}_E \cdot \nabla B$ is true for both clusters, but the sign change is only true for the lower-right cluster.

Standing out from the clusters, there is one non-thermal electron in Fig. \ref{fig:nonthermalpaths} that maintains a circular orbit and gains energy mainly from first-order Fermi acceleration. However, the acceleration is not very efficient as the relative energy gain is only 8\,\%. An efficient Fermi acceleration could be achieved closer to the O-type null point where the curvature is stronger. This requires that the electron has an initial position close to the null point, or a sufficient $\vec{E} \times \vec{B}$ drift towards it. On the other hand, orbits close to the null point -- where the characteristic magnetic field length is short and where the orbit size is comparable to the snapshot resolution -- are prone to numerical errors. The errors can produce slightly spiralling trajectories, and if the electron spirals towards the centre of the O-type null point, it experiences artificially strong first-order Fermi acceleration. For this reason, we filtered out the electrons that have their final position within a box of size 12.7 km $\times$ 4.5 km and 4.5 km $\times$ 4.5 km around the O-type null points associated with the upper and lower plasmoid, respectively. Of the 13 million electrons in the ensemble, 17 thousand end up in these boxes.

In general, all non-thermal electrons are mainly accelerated from initial positions in the current sheet region. Here there are large gradients in the magnetic field and a significant electric field, which are required for both betatron and first-order Fermi acceleration (see Eqs. \eqref{eq:betatron} and \eqref{eq:fermi}). Figure \ref{fig:proposaldist} shows that the change in energy is particularly correlated with the larger upper-left plasmoid, which has started to merge with the reconnection exhaust. This plasmoid has weaker gradients than the rest of the current sheet, but has a stronger electric field. 

\begin{figure}
    \centering
    \includegraphics[width=1\linewidth]{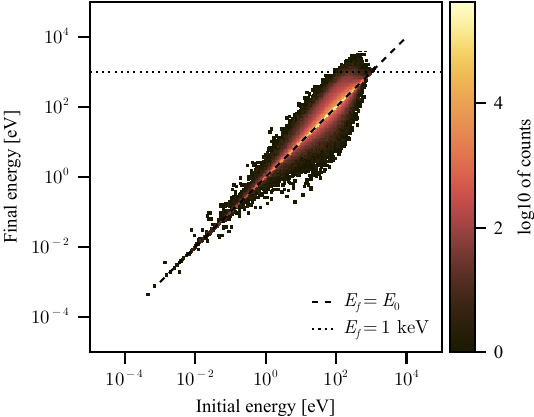} 
    \caption{Correlation between the initial ($E_0$) and final energy ($E_f$) of the electrons in the final test particle ensemble.}
    \label{fig:energycorrelation}
\end{figure}

When it comes to the electron acceleration as a function of initial energy, we know from Eq. \eqref{eq:gyrationenergy} that the energy change due to betatron acceleration is proportional to the magnetic moment of the particle, so that a high initial energy at low pitch angles increases the potential of strong betatron acceleration. And in Fig. \ref{fig:energycorrelation}, plotting the correlation between initial and final energy, we do see a tendency towards acceleration for high initial energies. All non-thermal electrons have an initial energy around or above 100 eV. 

\begin{figure}
    \centering
    \includegraphics[width=1\linewidth]{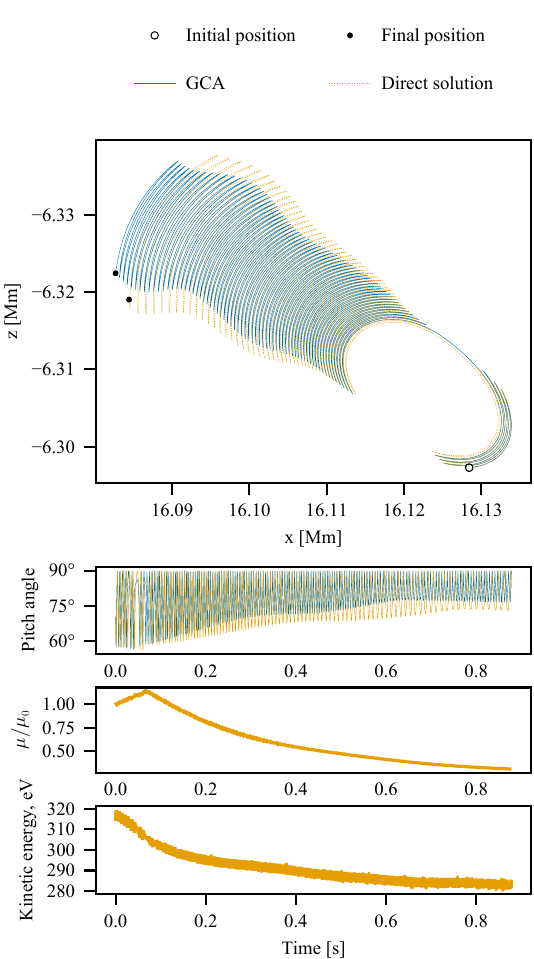}
    \caption{Comparison between the GCA and a direct solution to the Lorentz equation. The blue GCA trajectory is the same as in Fig. \ref{fig:trajectory}, and the orange trajectory is its equivalent full orbit. The lower panels show the pitch angle time evolution for both methods, as well as the time evolution of the magnetic moment (relative to the initial magnetic moment) and the kinetic energy for the direct solution. }
    \label{fig:directsolution}
\end{figure}

In Fig. \ref{fig:directsolution} we compare the non-thermal electron of Fig. \ref{fig:trajectory} with its equivalent direct solution of the Lorentz force. The comparison indicates that the betatron acceleration is, in this case, dependent on the conservation of magnetic moment. For all of the electrons investigated, a direct solution fails to reproduce the total energy increase achieved with the GCA. Although the trajectories are similar, the directly solved electrons lose magnetic moment, which decreases the average pitch angle and consequently increases the mirror length. The small reduction in kinetic energy may be first-order Fermi de-acceleration due to the curvature being anti-parallel to the $\mathbf{E} \times \mathbf{B}$ drift. This type of evolution is observed with several of the betatron-accelerated, non-thermal electrons, independent of the value of $r_L/L_B$. The energy gain in the GCA seems to be correlated with the variance of magnetic moment seen in the equivalent direct solution. For electrons that experience low $r_L/L_B$, the magnetic moment is closer to being conserved, and the energy gain is closer to zero. The electrons that experience a high $r_L/L_B$ gain more energy, but the magnetic moment is more variant. This correlation could explain the dependence of the non-thermal power law on the GCA tolerance (see Fig. \ref{fig:tolerancechange}).

\section{Conclusion}\label{section:conclusion}
We have developed a test particle code using the GCA and transported electrons in a static, 2D coronal, fan-spine configuration produced by the MHD code Bifrost. By using importance sampling in multiple stages, we were able to detect the formation of a non-thermal power law energy distribution from a statistically representative sample of electrons. Without the statistical weighing of the electrons, the energy fraction in the non-thermal power law increased by up to 2 orders of magnitude, indicating that the electron sampling technique has a significant impact when estimating non-thermal energy distributions using test particles. 

Furthermore, the shape of the non-thermal power law heavily depends on our GCA tolerance. This is partially because the GCA breaks down when the electrons get too close to a magnetic null point. A higher tolerance allowed us to simulate electrons with an initial position closer to the null points, and since betatron acceleration is proportional to the magnetic moment, which is inversely proportional to the initial magnetic field strength, it had a large effect on the number of non-thermal electrons. We also find a positive correlation between the tolerance and electron energy gain in general. Comparisons of the guiding centre trajectories with direct solutions to the Lorentz force indicated that the invariance of the magnetic moment was key to the betatron acceleration and that the energy gain was stronger for less adiabatic orbits. The sensitivity of the non-thermal power law to GCA tolerance, and the computational infeasibility of simulating particle ensembles with direct integration of the Lorentz force, suggests that a hybrid scheme is necessary to get a converging estimate of the true non-thermal power law in this static 2D MHD configuration. In addition to these conclusions, the following points should be considered in future work.

The particles were followed for 1\,s. The MHD timescales in and around the current sheet are of the same order. Assuming the electric and magnetic fields are static for 1\,s is therefore at the limit of what can be considered to be tolerable. A dynamic electromagnetic field may also help particles escape plasmoids, since the field lines are allowed to diffuse and reconnect over time.

The filtering of electrons close to O-type magnetic null points excludes potential non-thermal particles. These particles can be simulated more accurately using an interpolation scheme that ensures a divergence-free magnetic field, such as the one in \cite{mackayDivergencefreeMagneticField2006}. In addition, a hybrid scheme switching from GCA to full orbit integration is necessary when the particle gets too close to the magnetic null point. Non-adiabatic, O-type null point acceleration of the type discussed in \cite{zhaoMagneticIslandMerging2021} would then be possible, although only for very energetic electrons since for this type of acceleration, $r_L$ should be larger than the range of motion in the $xz$ plane, which means $r_L$ should be larger than the MHD grid size of 3.9 km. 

There is also the question of whether sustained, trapped, circular (or mirrored) motion around O-type null points will occur in more realistic 3D simulations. A third dimension will increase the complexity of the magnetic field, allowing the particles to escape through an additional dimension. A 3D version of the MHD simulation would also better facilitate the escape of accelerated electrons along the open field lines of the outer spine, and along the fan and inner spine towards the footpoints, as in \cite{pallister2021}. Moreover, 3D makes it easier for the particles to jump from one plasmoid or acceleration site to another, facilitating acceleration in consecutive stages.

\begin{acknowledgements}
This research is supported by the Research Council of Norway through its Centres of Excellence scheme, project number 262622. We thank Jan Karsten Trulsen for invaluable discussions on particle sampling. The simulation of test particle and data analysis were performed using the Julia programming language \citep{Julia-2017}. The Figures were produced using the open-source Julia-package Makie \citep{DanischKrumbiegel2021}. The authors thank the referee for valuable comments and suggestions.
\end{acknowledgements}

\bibliographystyle{aa}
\bibliography{references.bib} 

\begin{appendix}
\section{Rejection sampling}\label{appendix:rejectionsampling}
To sample initial particle positions from the electron density, which is a 2D inhomogeneous distribution always greater than zero (here called the target distribution $f(\vec{x})$), we used rejection sampling (see e.g. Sect. 7.3.6 in \citealt{numericalrecipes}). We drew our samples $\vec{x}_i$ from a uniform probability density distribution with limits identical to the target distribution, and accept the samples with a probability proportional to $f(\vec{x}_i)$. In this way, our accepted draws will reflect the shape of the target distribution, at the cost of rejected draws. The algorithm goes as follows:
\begin{enumerate}
    \item Draw a random number $u$ from the uniform probability distribution $U (x_a,x_b)$. The limits $x_a$ and $x_b$ must coincide with the $x$-limits of the target distribution $f=f(x, y)$.
    \item Draw another random number $v \in U (y_a,y_b)$, where the limits $y_a$ and $y_b$ coincide with the $y$-limits of $f$.
    \item Draw a third random number $w \in U(0,c)$, where the constant $c$ is equal to or greater than the maximal value of $f(x, y)$.
    \item If $w \leq f(u, v)$ accept the sample $(u, v)$. Otherwise, reject $(u, v)$.
    \item Repeat the above steps until you have the desired number of accepted samples.
\end{enumerate}
The ratio of accepted draws to rejected draws strongly depend on the shape of the target distribution. For an optimal choice of $c$, the acceptance ratio approaches 1 as the target distribution approaches uniform. At the other extreme, the acceptance rate would be zero for a Dirac delta function. To approximate the target distribution from the drawn samples, we had to bin the samples in a histogram. The accuracy of each bin is inversely proportional to the square root of the number of samples in that bin.

\section{Importance sampling}\label{appendix:importancesampling}
\begin{figure}
    \resizebox{\hsize}{!}{\includegraphics{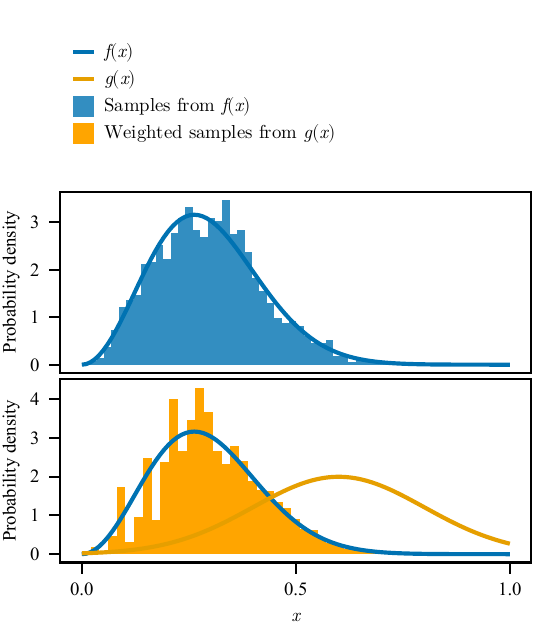}}
    \caption{
    Showcase of importance sampling. The upper panel shows a
    histogram of 2000 samples directly sampled from a Maxwell-Boltzmann
    probability density distribution, $f(x)$. In the lower panel, the same amount of samples are sampled from a Gaussian probability density distribution, $g(x)$, which is shifted towards higher $x$ values compared to the Maxwell-Boltzmann distribution. This shift increases the fraction of samples with high $x$-values compared to the direct sampling, and if the samples are weighted with the ratio $f(x)/g(x)$ when binned, the resulting histogram will estimate the Maxwellian-probability distribution, but this time with higher statistical accuracy for the higher values of $x$.
    }
    \label{fig:importancesampling}
\end{figure}

In importance sampling (see e.g. Sect. 7.9.1 in \citealt{numericalrecipes}) we draw samples from a selected proposal distribution instead of the target distribution. To recreate the target distribution, we multiply the samples with importance weights. The more probable the sample is in the target distribution, the higher the weight, and the more probable the sample is in the proposal distribution, the lower the weight. The proposal distribution is required to cover the same domain as the target distribution, but is otherwise arbitrary, only limited by the feasibility of sampling from it. The benefit of importance sampling is two-folded. First, the proposal distribution may be easier or faster to sample from (e.g. a uniform distribution). Second, biasing the sampling towards values that otherwise would be rarely sampled allows us to increase the statistical accuracy of the target distribution at those locations. As an example, Fig. \ref{fig:importancesampling} displays importance sampling of a Maxwell-Boltzmann probability density distribution. Direct sampling will estimate the high-value tail of the distribution with poor accuracy, since this part of the distribution is rare. A well-placed proposal distribution, like a Gaussian shifted towards the tail, will significantly increase the accuracy of the high-value estimates, at the cost of the accuracy of the lower value estimates.

\subsection{Calculating the importance weights}\label{section:importanceweights}
Consider the arbitrary function $h(x)$ and the probability density distribution $f(x)$. The integral 
\begin{equation}
    \mathcal{I} = \int h(x)f(x)\dd x = E^f\left[h(\mathrm{x})\right]
\end{equation}
is the expectation value of $h(x)$ in $f$, where $\mathrm{x}$ is a random variable that is distributed according to $f(x)$. The integral can be approximated by the sample mean,
\begin{equation}
    \mathcal{I} = E^{f}\left[h(\mathrm{x})\right] \approx \frac{1}
    {n} \sum^n_i h(x_i),
\end{equation}
where $x_1, \ldots, x_n$ are realisations (samples) of $\mathrm{x}$, and $n$ is 
the number of these realisations. Next, we introduce the proposal probability density 
distribution $g(x)$ into the integral,
\begin{equation}
    \mathcal{I} = \int h(x)f(x)\frac{g(x)}{g(x)} \dd x = E^{g}\left[h(\mathrm{x})\frac{f(\mathrm{x})}{g(\mathrm{x})}\right].
\end{equation}
In this case, $\mathcal{I}$ is formulated as the expectation value of $h(x)f(x)/g(x)$ in $g$, and $\mathrm{x}$ is a random variable distributed according to $g(x)$. We can then approximate the integral with a different sample mean, namely
\begin{equation}
    \mathcal{I} \approx \frac{1}{m} \sum^m_j h(x_j)\frac{f(x_j)}{g(x_j)},
\end{equation}
where $x_1, \ldots, x_m$ are realisations of the random variable $\mathrm{x}$, distributed according to $g(x)$. The ratio $f(x_j)/g(x_j)$ is the importance weight of the sample $h(x_j)$. In the example in Fig. \ref{fig:importancesampling}, $h(x)=x$, $f(x)$ is the Maxwell-Boltzmann target distribution, and $g(x)$ is the Gaussian proposal distribution.

\end{appendix}

\end{document}